\documentclass[12pt]{article}
\usepackage{graphicx}

\newcommand\pubnumber{CMS CR-2017/390}  
\newcommand\pubdate{\today}

\def\institute{Department of Physics and Astronomy\\
University of Bologna and INFN, I-40127 Bologna, ITALY}

\def\Title#1{\begin{center} {\Large #1 } \end{center}}
\def\Author#1{\begin{center}{ \sc #1} \end{center}}
\def\Address#1{\begin{center}{ \it #1} \end{center}}

\newcommand\pubblock{\rightline{\begin{tabular}{l} \pubnumber\\
         \pubdate  \end{tabular}}}
\newenvironment{Abstract}{\begin{quotation}  }{\end{quotation}}
\newenvironment{Presented}{\begin{quotation} \begin{center} 
             PRESENTED AT\end{center}\bigskip 
      \begin{center}\begin{large}}{\end{large}\end{center} \end{quotation}}

\def\beq{\begin{equation}}
\def\eeq#1{\label{#1}\end{equation}}
\def\eeqn{\end{equation}}

\def\beqa{\begin{eqnarray}}
\def\eeqa#1{\label{#1}\end{eqnarray}}
\def\eeqan{\end{eqnarray}}

\let\bar=\overbar

\def\Dslash{\not{\hbox{\kern-4pt $D$}}}
\def\dslash{\not{\hbox{\kern-2pt $\del$}}}

\def\msb{{\bar{\ssstyle M \kern -1pt S}}}

\begin{document}
\begin{titlepage}
\pubblock

\vfill

\Title{Recent Top Quark Mass Measurements from CMS}

\vfill

\Author{\textsc{Andrea Castro, on behalf of the CMS Collaboration}}
\Address{\institute}
\vfill

\begin{Abstract}
A variety of top quark mass measurements has been made in the recent years by the CMS Collaboration. 
The most recent measurements performed at 8 TeV are reported here, along with a new measurement based on data collected in 2016 at 13 TeV. The current combination of these measurements has  a relative uncertainty smaller than 0.3\%, making the top quark the most accurately measured quark.
\end{Abstract}
\vfill
\begin{Presented}
$10^{th}$ International Workshop on Top Quark Physics\\
Braga, Portugal,  September 17--22, 2017
\end{Presented}
\vfill
\end{titlepage}
\def\thefootnote{\fnsymbol{footnote}}
\setcounter{footnote}{0}

\section{Introduction}
 At the CERN LHC, the strong production of top quark-antiquark ($t\bar t$) pairs in proton-proton ($pp$) collisions  is quite abundant: more than 5 million $t\bar t$ pairs have been produced in the Run 1 at $\sqrt{s}=7$ and 
8 TeV. The events collected have a  $t\bar t\to W^+bW^-\bar b$ final state distinguished by the $W$'s decay into {\it dilepton}, {\it lepton+jets}, and {\it 
all-jets} channels. About a million more top quarks have been produced singly.
\par
Since the top quark mass $M_t$ is a free parameter of the Standard Model (SM) it is important to measure it directly from the reconstruction of its decay products or, indirectly, comparing the $t\bar t$ cross section to theoretical expectations, yielding a top quark mass determined in a well-defined theoretical scheme as used in perturbative QCD calculations, e.g. the so-called {\em pole mass}. 
Precise measurements of $M_t$, the $W$ mass $M_W$, and the Higgs boson mass $M_H$ are crucial, and can be used to test the self-consistency of the SM~\cite{sm-global-fit}, or to explore models for new physics~\cite{bsm-ref}.
In addition, $M_t$ and $M_H$ are related to the vacuum stability~\cite{vacuum-stability} of the SM, with the current value $M_H\approx 125$ GeV~\cite{pdg} corresponding to a near-criticality.
\par

\section{Measuring the top quark mass at CMS}
The $t\bar t$ (or single top quark) events collected by CMS~\cite{cms-ref} in $pp$ collisions have common physics 
signatures: isolated 
leptons ($\ell=e$ or $\mu$) with high transverse momentum ($p_{\rm T}$); high-$p_{\rm T}$ jets, some of which associated to the hadronization of $b$ quark (i.e. $b$-jets); missing transverse momentum, $p_{\rm T}^{\rm miss}$,  associated to neutrinos. Because of the large $\sqrt{s}$ value, top quarks can be produced at  high $p_{\rm T}$, originating the so-called  {\em boosted jets}.
The reconstruction of the $pp\to t\bar t\to W^+bW^-\bar b$ final state (or the corresponding one for single top quark events) is based on these physics objects, but ambiguities and permutations need to be considered when mapping them to the leptons/quarks of the final state. Furthermore, there is an additional uncertainty associated to the knowledge of the absolute value of jet energies, i.e. the so-called {\em jet energy scale} (JES), and the $p_{\rm T}^{\rm miss}$ sharing between multiple neutrinos.
\par
Given the large number of top quarks produced, the statistical uncertainties are typically small, while the systematic ones dominate. Sources of uncertainty related to experimental effects, signal and background modeling are studied in detail.
\par
We present in the following the most recent measurements of $M_t$ with $19.7$ fb$^{-1}$ of integrated luminosity collected in $pp$ collisions at $\sqrt{s}=8$ TeV.  A new measurement, conducted with $35.9$ fb$^{-1}$ of $pp$ collisions at $\sqrt{s}=13$ TeV, is also reported.

\par
\subsection{ Single top quark, $\mu$+jets}
For this channel, we select events with one muon and two jets, one of which $b$-tagged, and recur to a  template method~\cite{cms-single-ref}, using the invariant mass $m_{\mu\nu b}$ of the muon, 
 the $b$-jet, and the neutrino whose momentum is inferred constraining the $\mu\nu$ invariant mass to $M_W$. 
The  $m_{\mu\nu b}$  distribution  is described by analytical functions whose parameters are related to $M_t$. The template fit, see Fig.~\ref{mass-cms} (left),  returns a value for $M_t$ of
 $172.95\pm 0.77\,({\rm stat})\, ^{+0.97}_{-0.93}\,({\rm syst})$ GeV, with a total uncertainty of 1.24 GeV ($0.72\%$).
 The systematic uncertainty is dominated by  the  JES uncertainty (0.68 GeV), the background modeling  (0.39 GeV), and the fit calibration (0.39 GeV). 
\subsection{Boosted top quark}
\begin{figure}[htb]
\begin{tabular}{ccc}
~~~~ & 
\begin{minipage}{0.4\textwidth}
\includegraphics[width=6.5cm]{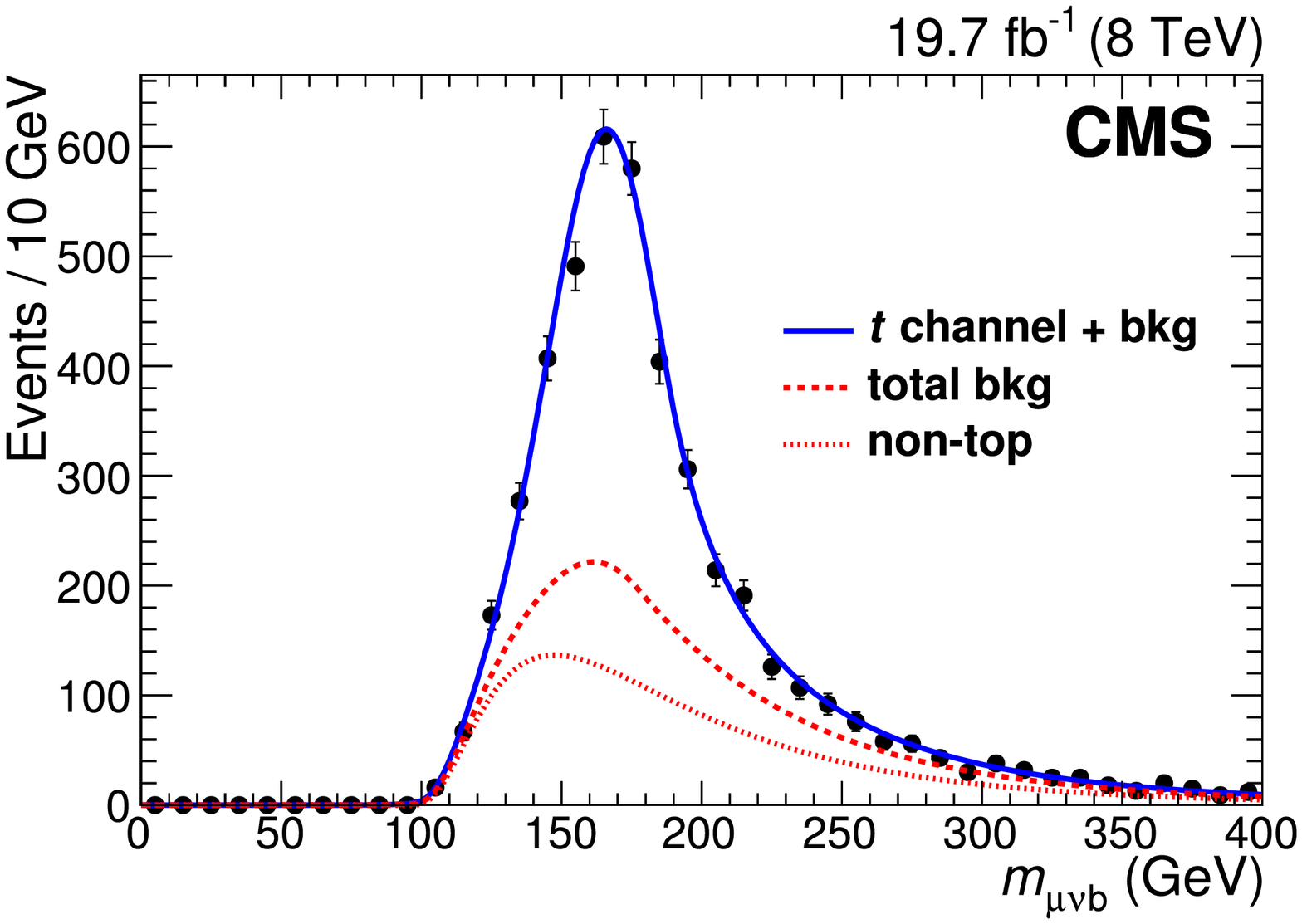}
\end{minipage}
& 
\begin{minipage}{0.4\textwidth} 
\includegraphics[width=7.0cm]{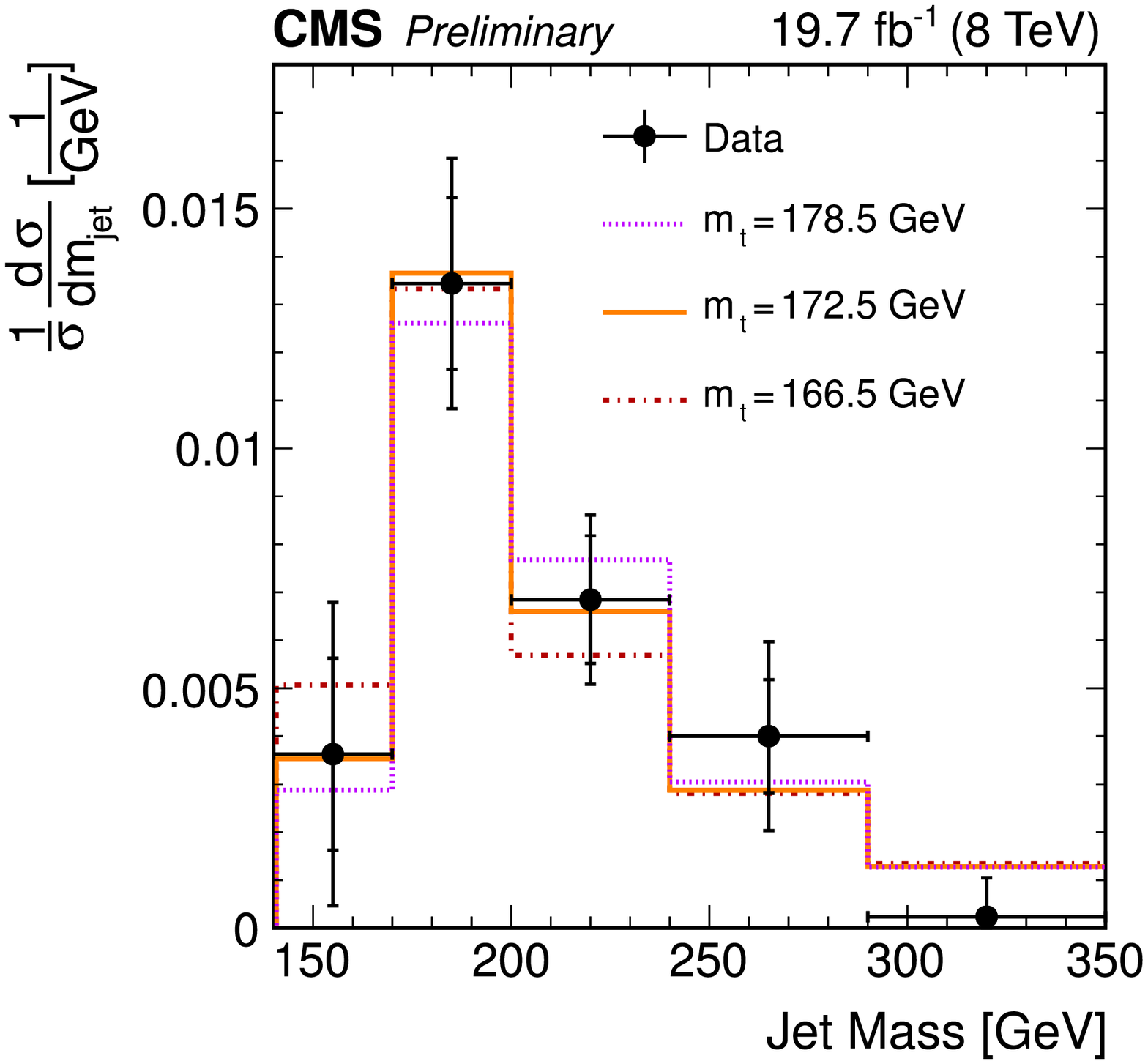}
\end{minipage}
\end{tabular}
\caption{ CMS top quark events at $\sqrt{s}=8$ TeV. Left: $m_{\mu\nu b}$ fit for single top quark, $\mu$ + jets events~\cite{cms-single-ref}. Right:  normalized $m_{\rm jet}$ differential cross section for boosted top quark events~\cite{cms-boost-ref}. }
\label{mass-cms}
\end{figure}

In this case, we consider events with one lepton ($e$ or $\mu$), at least two wide (boosted) jets, at least three narrow jets, including at least one $b$-jet.
The analysis~\cite{cms-boost-ref} is intended to  measure the differential cross section as a function of the boosted jet invariant mass  $m_{\rm jet}$, for jets with $p_{\rm T}>500$ GeV. The normalized differential cross section depends indeed on $M_t$, as shown in  Fig.~\ref{mass-cms} (right), so from it we can infer the value of $M_t$ which best reproduce the data. A template fit returns a value $M_t=170.8\pm 6.0\,({\rm stat})\pm 2.8\,({\rm syst})\pm 4.6\,({\rm model})\pm 4.0\,({\rm theory}) $ GeV, with a total uncertainty of 9.0 GeV ($5.3\%$),  dominated by contributions from the signal modeling.

\par\noindent

\subsection{Alternative methods}
The ``standard'' measurements described above are strongly affected by the JES and hadronization modeling. To reduce these systematic uncertainties we can use cleaner observables, i.e. avoid using jets. 
The top quark mass needs then to be reconstructed from ``alternative'' observables, like for instance the invariant mass of $\ell$ plus a
 reconstructed $J/\psi$~\cite{cms-alt-jpsi-ref}, or the invariant mass of the system made of $\ell$ plus a secondary vertex~\cite{cms-alt-vtx-ref}. Additional measurements are based on $M_t$-sensitive observables like the invariant masses of the $b\ell$ and $b\ell\nu$ systems, or the so-called {\em stransverse mass} of the $bb$ pair~\cite{cms-alt-maos-ref}. The value of $M_t$ obtained in the last case amounts to 
$M_t=172.22\pm 0.18\,({\rm stat})\,^{+0.89}_{-0.93}\,({\rm syst})$ GeV, with systematic uncertainties  dominated by contributions from JES, $b$-fragmentation and top quark $p_T$  modeling.
\par
A more clearly defined value for  $M_t$  can be derived from the comparison of the $t\bar t$ production cross section to theoretical predictions. Such indirect measurement~\cite{cms-alt-pole-ref} leads to a pole-mass value of  $173.8\, ^{+1.7}_{-1.8}$ GeV, with dominant systematic uncertainties related to parton distribution functions and integrated luminosity.
\par
\subsection{Run 1 top quark mass combinations}
\begin{figure}[htb]
\begin{tabular}{ccc}
~~~ & 
\begin{minipage}{0.4\textwidth}
\includegraphics[width=6.5cm]{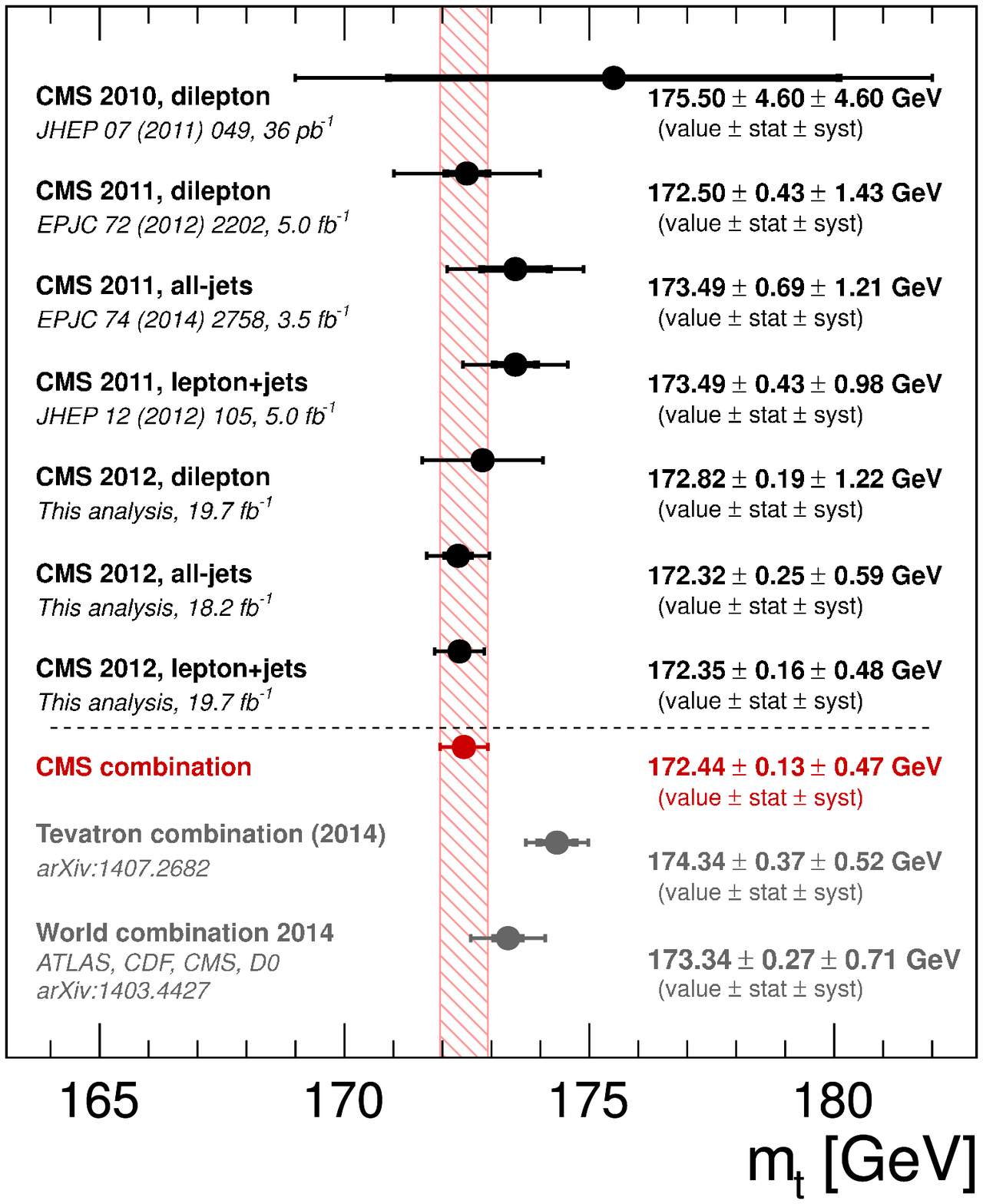}
\end{minipage}
  &  ~~~
\begin{minipage}{0.4\textwidth} 
\includegraphics[width=6.5cm]{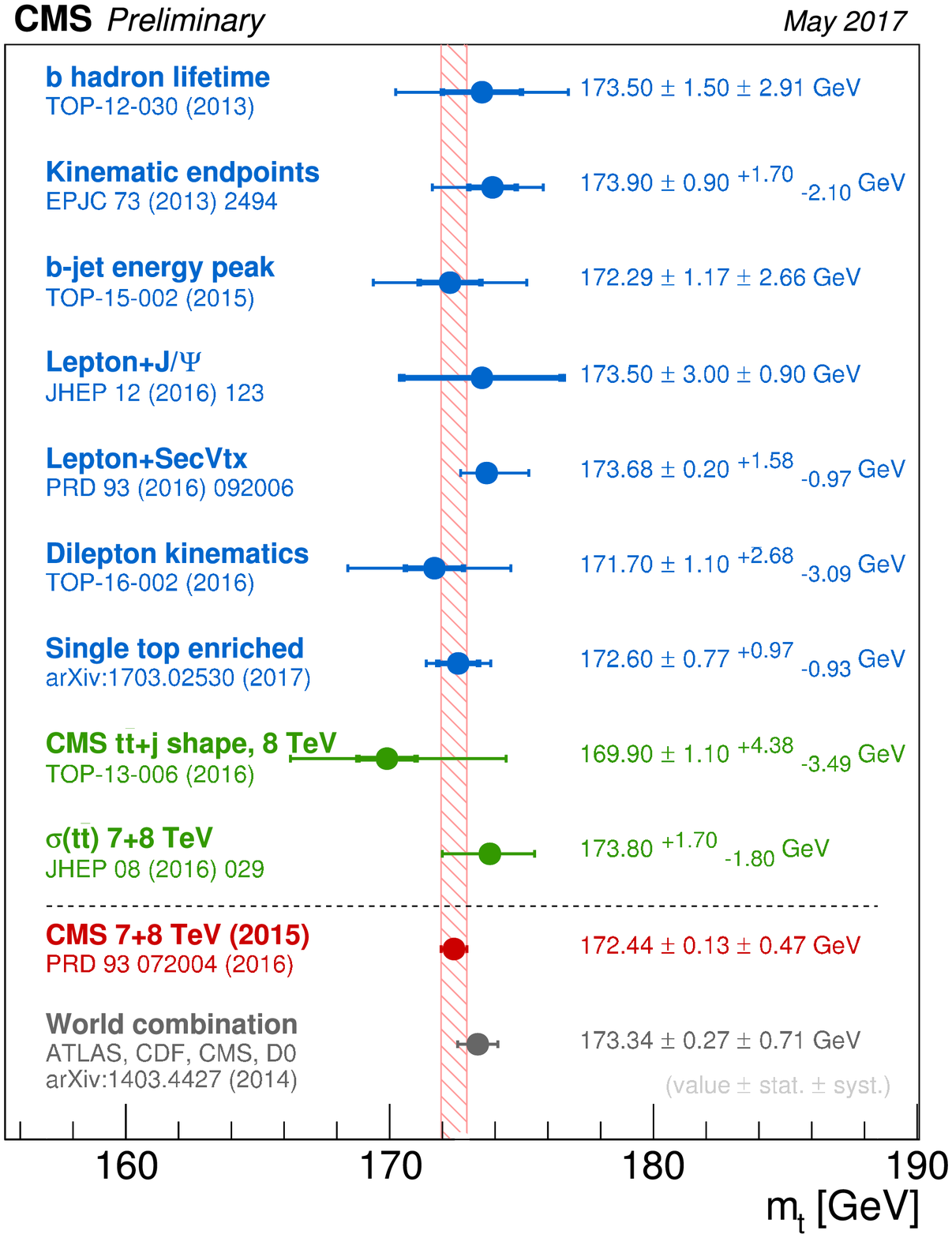}
\end{minipage}
\end{tabular}
\caption{Summary of CMS $M_t$ measurements at 7 and 8 TeV. Left: ``standard''~\cite{cms-ave-ref} measurements. Right: ``alternative''~\cite{cms-alt-ave-ref} measurements.}
\label{mass-combos}
\end{figure}

The mass measurements performed by CMS at 7 and 8 TeV, summarized in Fig.~\ref{mass-combos} (left), are combined~\cite{cms-ave-ref}  yielding a value  $M_t=172.44\pm 0.13\,({\rm stat})\pm 0.47\,({\rm syst})$ GeV, with a total uncertainty of 0.49 GeV corresponding to 0.28\%. A similar combination has been calculated for the alternative measurements~\cite{cms-alt-ave-ref}, see Fig.~\ref{mass-combos} (right), yielding a value $M_t=172.58\pm 0.21\,({\rm stat})\,\pm 0.72\,({\rm syst})$ GeV, slightly less precise, but providing an independent verification with different systematic uncertainties. 
\par\noindent
These combinations, made individually by CMS, are consistent with those performed by the ATLAS Collaboration~\cite{atlas-ave-ref}, and  are more precise than the 2014 world average~\cite{world-ave-ref} $M_t=173.34\pm 0.76$ GeV, which had a  total uncertainty of $0.44\%$.

\subsection{Lepton + jets at 13 TeV}
\begin{figure}[htb]
\begin{tabular}{ccc}
~~~ & 
\begin{minipage}{0.4\textwidth}
\includegraphics[width=6.5cm]{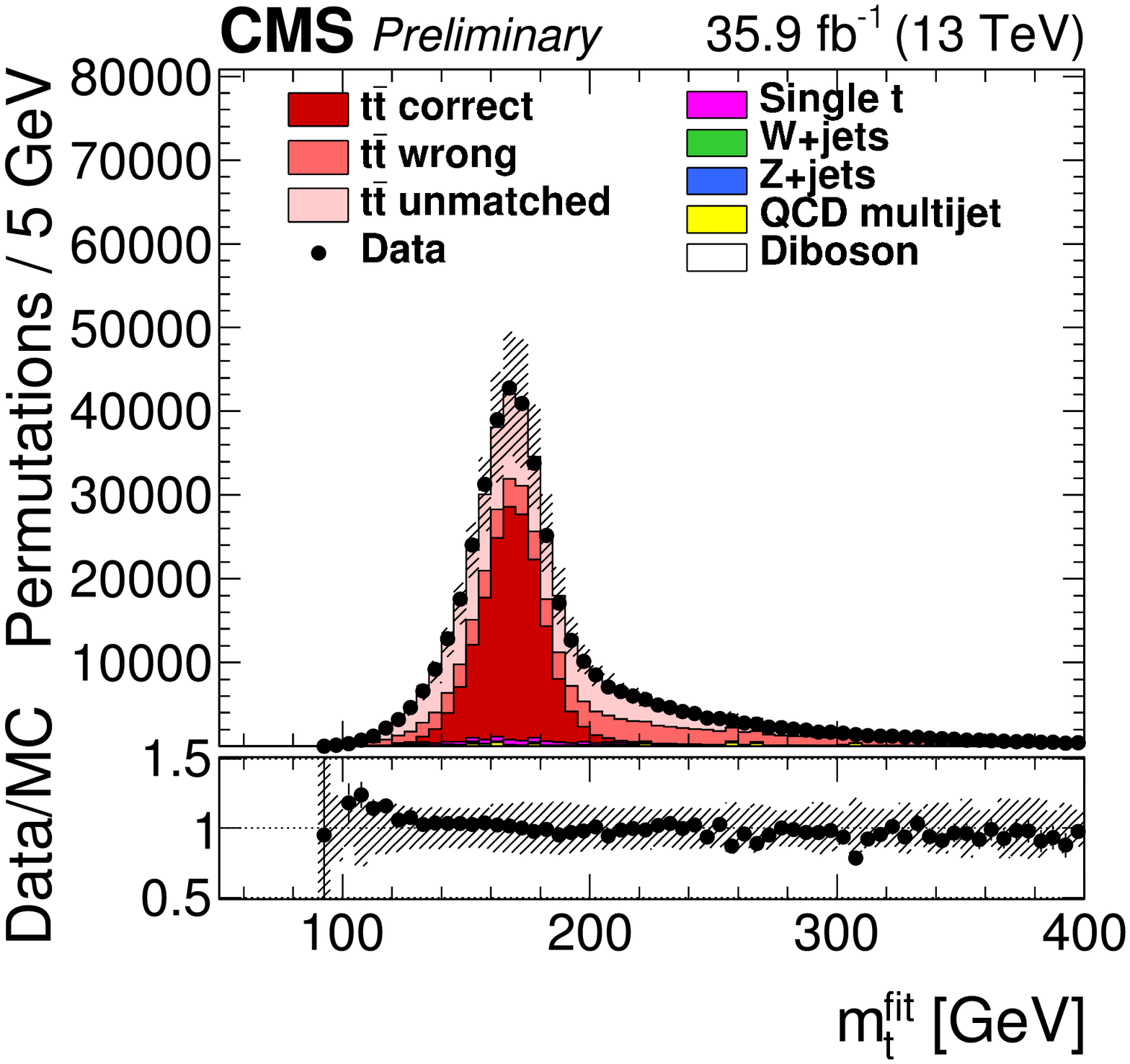}
\end{minipage}
& ~~~
\begin{minipage}{0.4\textwidth} 
\includegraphics[width=6.5cm]{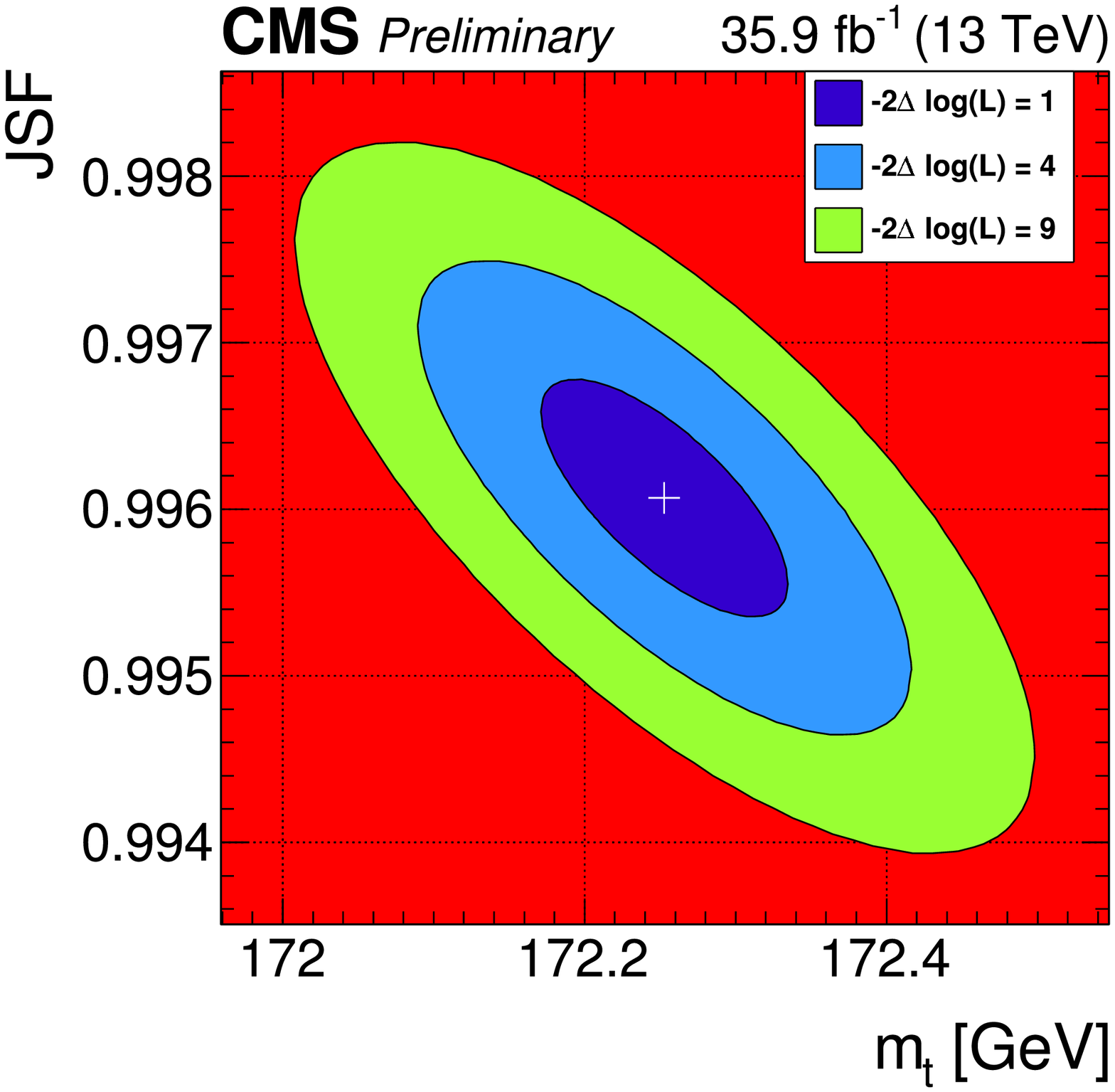}
\end{minipage}
\end{tabular}
\caption{CMS lepton + jets events at 13 TeV~\cite{cms-13-ref}. Left: distribution of  $m_t^{\rm fit}$. Right: 2-D contour plot $JSF$ vs $M_t$.}
\label{mass-cms-13}
\end{figure}

For this very recent measurement~\cite{cms-13-ref}, events are selected with one $\mu$,  and at least four jets, including two $b$-jets.
 The measurement is based on the so-called {\em ideogram method}.
 Starting from the kinematical reconstruction of the $WbWb$ final state, an event likelihood is computed as a function of $M_t$ and of the reconstructed $W$ boson mass, convoluting Breit--Wigner (or similar) distributions with experimental resolutions. Multiple combinations for the jet-to-quark matching are considered with weights depending on the goodness of the kinematic fit. The signal purity is quite improved by a request on the fit probability, yielding a very peaked distribution for the reconstructed top quark mass $m_t^{\rm fit}$, as shown in Fig.~\ref{mass-cms-13} (left). A likelihood fit, based on distributions of the $W$ boson and top quark reconstructed masses,  allows  an in situ calibration of  a factor $JSF$ which modifies the default JES, while considering the expected uncertainty on $JSF$ (``hybrid'' fit).
The values returned by the fit, see Fig.~\ref{mass-cms-13} (right), 
are $JSF=0.996\pm 0.008$ and  $M_t=172.25\pm 0.08\,({\rm stat}+JSF)\pm 0.62\,({\rm syst})$ GeV, with a total uncertainty of 0.63 GeV ($0.36\%$).
 The systematic uncertainty is  dominated by the flavor-dependence of the JES (0.39 GeV),  color reconnection effects effects (0.31 GeV), and the choice of the matrix-element generator (0.22 GeV).

\section{Summary}
Since the  top quark discovery, made 22 years ago, the measurement of its mass has been actively pursued recurring to a variety of channels and techniques.
The precision reached is quite impressive, smaller than $0.3\%$, and expected to improve with ongoing and future measurements at the LHC, including refinements in the methodology.  
To  shed light on fundamental  cosmological issues and on physics beyond the SM, it will be important to reduce the systematic uncertainties, mainly those related to signal modeling, with a thorough  tuning of the parameters in the Monte Carlo generators.

\end{document}